\documentclass[a4paper,12pt]{article}
\usepackage{amsfonts,amsmath,amsthm,amssymb,amscd}
\usepackage{newpxtext,newpxmath} 
\usepackage[square,numbers]{natbib} 
\usepackage{bibspacing} 
\usepackage[nobottomtitles*]{titlesec}
\titleformat{\section}{\Large\bf}{\thesection}{.8em}{}
\titlespacing*{\section}{0pt}{*3}{.9ex}[\fill]
 
\usepackage{pict2e} 
\sfcode`P=1000  
\usepackage{microtype} 
\usepackage{graphicx}
\usepackage[colorlinks,linkcolor=blue,citecolor=blue]{hyperref} 

\usepackage{xcolor}
\newtheorem{theorem}{Theorem}
\newtheorem{lemma}{Lemma}
\newtheorem{proposition}{Proposition}
\theoremstyle{definition}

\newcommand{\supp}{\textup{\small\textsf{supp}}} 
 
\newcommand{\RS}{\textup{\small\textsf{rowspace}}} 
\newcommand{\NS}{\textup{\small\textsf{nullspace}}} 
\newcommand{\twovec}[2]{\small\Bigl(\begin{matrix}#1\\[-.7ex]#2\end{matrix}\Bigr)}
\newcommand{\reals}{\mathbb{R}} 
\newcommand{\eps}{\varepsilon} 
\newcommand{\T}{^{\top}} 
\newcommand{\0}{{\mathbf0}}
\newcommand{\1}{{\mathbf1}}
\def\2{\textup{\small\textsf{GF}}(2)}
\def\myproofof#1{\noindent{\em Proof of #1.\enspace}}
\def\myproof{\noindent{\textit{Proof.\enspace}}}
\def\endproof{\hfill\strut\nobreak\hfill\tombstone\par\smallbreak}
\def\tombstone{\hbox{\lower.4pt\vbox{\hrule\hbox{\vrule
  \kern7.6pt\vrule height7.6pt}\hrule}\kern.5pt}} 
\newcommand{\maxi}{\mathop{\hbox{\textup{maximize }}}}
\newcommand{\mini}{\mathop{\hbox{\textup{minimize }}}} 
\newcommand{\subj}{\hbox{\textup{subject to ~ }}} 

\newdimen\einr\einr1.8em
\newdimen\rmeinr\rmeinr1.8em
\newdimen\tmp 
\newcommand{\abs}[1]{\par\hangafter=1\hangindent=\einr
  \noindent\hbox to\einr{#1\hfill}\ignorespaces} 
\newcommand\rmitem[1]{\abs{\textup{#1}}} 
\newcommand\bullitem{\tmp\einr\einr\rmeinr\abs{\raise.17ex\hbox{\kern7pt\scriptsize$\bullet$}}\einr\tmp}

\parindent\einr
\title{Zero-Sum Games and Linear Programming Duality}
\author{Bernhard von Stengel\\
\small
Department of Mathematics, London School of
Economics, \sf b.von-stengel@lse.ac.uk}
\date{\normalsize October 23, 2023}

\begin{document}
\maketitle

\begin{abstract}
\noindent
The minimax theorem for zero-sum games is easily proved from
the strong duality theorem of linear programming.
For the converse direction, the standard proof by Dantzig
(1951) is known to be incomplete.
We explain and combine classical theorems about solving
linear equations with nonnegative variables to give a
correct alternative proof, more directly than Adler (2013).
We also extend Dantzig's game so that any max-min strategy
gives either an optimal LP solution or shows that none
exists.

\end{abstract}

\arraycolsep.2em


\section{Introduction and summary}

LP duality (the strong duality theorem of linear
programming) is a central result in optimization.
It helps proving many results with ease, such as the minimax
theorem for zero-sum games, first proved by von Neumann in
1928 \cite{vN1928}.
In October 1947, George Dantzig explained his nascent ideas on linear
programming to John von Neumann \cite[p.~45]{dantzig1982}.
In response, he got an ``eye-popping'' lecture on LP
duality, which von Neumann conjectured to be equivalent to
his minimax theorem.
This ``equivalence'' is commonly assumed (for example,
Schrijver \cite[p.~218]{Schrijver}), but on closer
inspection does not hold at all.

``Equivalence'' is actually not a good term -- all
theorems, as logical statements without free variables, are
equivalent, to ``true''.
We therefore say that theorem~A \textit{proves} (rather than
``implies'') theorem~B, typically by a suitable but different
use of the variables in theorem~A, and state straightforward
proof relations of this kind as propositions (see
Proposition~\ref{p-farkas} for an example).


The classic proof by Dantzig \cite{dantzig1951} of LP duality from
the minimax theorem needs an additional assumption
about the game solution, namely strict complementarity in
the last column of the game matrix  that corresponds to the
right-hand side of the LPs.
(We state Dantzig's game in (\ref{B}) below; it differs from
the original in a trivial change of signs so that the primal
LP is a maximization problem subject to upper bounds, in
line with the row player in a zero-sum game as the
maximizer.)
This complementarity assumption, acknowledged by 
Dantzig \cite{dantzig1951}\cite[p.~291]{Dantzig},
applies only to non-generic LPs and seems technical.
Adler \cite{adler2013} fixed this ``hole'' in Dantzig's
proof, and showed how an algorithm that solves a zero-sum
game can be used to either solve an LP or certify that it
has no optimal solution.
Recently, Brooks and Reny \cite{BReny} gave a zero-sum game
whose solution directly provides such a solution or
certificate.

The aim of this article is to clarify the underlying
problem, with two new main results (explained later).
Our narrative is self-contained, not least because LP
duality is so familiar that it can be overlooked as a silent
assumption.
For example, reducing optimality of maximizing $c\T x$
subject to $Ax\le b$, $x\ge\0$ to feasibility of $Ax\le b$,
$x\ge\0$, $A\T y\ge c$, $y\ge\0$, $ b\T y\le c\T x$
assumes that there cannot be a positive ``duality gap''
$b\T y-c\T x$, which is the strong duality theorem.
Our presentation shows how one could prove, in full, LP
duality via the minimax theorem, if one were to take that
route.
Some of the presented less-known elegant proofs from the
literature are also of historical interest.

Dantzig's assumption holds if a pure strategy that is a best
response in every solution of the zero-sum game has positive
probability in some solution.
As noted by Adler \cite[p.~167]{adler2013}, this can be
shown (e.g., \cite[p.~742]{raghavan1994}) using a version of
the Lemma of Farkas \cite{farkas1902}.
However, the Lemma of Farkas proves LP duality directly.
Our first, easy observation is that Dantzig's assumption
amounts to the Lemma of Tucker \cite{tucker1956}.
This, in turn, directly proves the Lemma of Farkas
\cite[p.~7]{tucker1956}, even for the special case
of Dantzig's game (Proposition~\ref{p-Bfa} below).
The assumption is therefore extremely strong and in a sense
useless for proving LP duality from the minimax theorem.
Curiously, Tucker did not consider the converse that in nearly
the same way the Lemma of Farkas proves his Lemma (see
Proposition~\ref{p-tufa} below).
This suggests that Tucker thought he had proved a more general
statement.
Tucker's proof of his Lemma is indeed short and novel, but
in this light we agree with Adler's view of Tucker's Lemma
as a ``variant of Farkas's Lemma'' \cite[p.~174]{adler2013}.

LP duality and the minimax theorem are closely related to
solving, respectively, inhomogeneous and homogeneous linear
equations in nonnegative variables.
The Lemma of Farkas characterizes when the inhomogeneous
linear equations $Ax=b$ have no solution vector $x$ such
that $x\ge\0$.
The Theorem of \textit{Gordan} \cite{gordan1873}
characterizes when the homogeneous equations $Ax=\0$ have no
solution $x\ge\0$ other than the trivial one $x=\0$.
Gordan's Theorem and its ``inequality version'' due to
Ville \cite{ville1938} prove the minimax theorem and vice versa.

Our first main result, Theorem~\ref{t-gotu} in
Section~\ref{s-gotu}, is a proper proof of LP duality from
the minimax theorem.
Inspired by Adler \cite[section~4]{adler2013}, we use
Gordan's Theorem to prove the \textit{Theorem} of Tucker
\cite{tucker1956}, an easy but powerful generalization of
his Lemma (like Broyden \cite{broyden2001} we think that it
deserves more recognition).
Tucker's Theorem shows that any system of homogeneous
equations $Ax=\0$ such that $x\ge\0$ has a natural partition
of its solution vector~$x$ into a set of variables that can
take positive values and the others that are 
zero in any nonnegative solution.
It is easy to see that one can drop the nonnegativity
requirement for the variables that can be positive. 
By \textit{eliminating} these unconstrained variables from
the system $Ax=\0$ with a bit of linear algebra, applying 
Gordan's Theorem to the variables that are always zero
in any nonnegative solution then gives Tucker's Theorem.
Compared to the detailed computations of this variable
elimination by Adler \cite{adler2013}, our proof is
self-contained and more direct.
Using Dantzig's game (\ref{B}), Tucker's theorem proves LP
duality in a stronger version, namely the existence of a
``strictly complementary'' solution to the LPs if they are
feasible (Proposition~\ref{p-strict} below).

Our second main result, Theorem~\ref{t-BM} in
Section~\ref{s-M}, extends Dantzig's elegant game (\ref{B})
with an extra row in (\ref{BM}) that ``enforces'' the desired
complementarity in the last column.
\textit{Every} max-min strategy of this game either gives an
optimal pair of solutions to the primal and dual LPs, or 
represents an unbounded ray for at least one of the LPs if
it is feasible, so that the other LP is therefore infeasible.
This result is similar to Adler's ``Karp-type'' reduction of
an LP to a zero-sum game \cite[section~3.1]{adler2013}, but
with the extra certificate of infeasibility.
It is also similar to, and inspired by, the main result of
Brooks and Reny \cite{BReny}.
The proof of Theorem~\ref{t-BM} (in a separate
Theorem~\ref{t-DM}) does \textit{not} rely on LP duality and
was surprisingly hard to find. 
Compared to either \cite{adler2013} or \cite{BReny}, our
game (\ref{BM}) more naturally extends Dantzig's original game.
Similar to both, it imposes an upper bound on the LP
variables that does not affect whether the LPs are feasible.
This bound follows from Carath\'eodory's theorem
\cite{caratheodory1911} that nonnegative solutions $x$ to
$Ax=b$ can be found using only linearly independent columns
of~$A$ (of which there are only finitely many sets).
That bound is determined apriori and of polynomial encoding
size from the sizes of the entries of $A$ and $b$ if these
are integer or algebraic numbers, otherwise abstractly from
all ``basic solutions'' $x$ to $Ax=b$.

%
We give a self-contained introduction to linear programming
duality (for LPs in inequality form) and to the minimax
theorem in Section~\ref{s-state}.
Section~\ref{s-farkas} recalls how LP duality is proved from
the Lemma of Farkas. 
The theorems of Gordan \cite{gordan1873} and Ville
\cite{ville1938} are the topic of Section~\ref{s-govi}.
Stiemke \cite{stiemke1915} gave a two-page proof of the Theorem
of Gordan (without referencing it, even though
published in the same journal, presumably with no editor
around to remember it).
His proof uses implicitly that the null space and row space
of a matrix are orthogonal complements.
But there are no matrices in these papers -- people
manipulated linear equations with their unknowns instead.
For historical interest, and because of its structural
similarity to Tucker's proof of his Lemma \cite[p.~5--7]{tucker1956},
we reproduce Stiemke's proof
in Section~\ref{s-sti}.
We also present a most elegant half-page proof of the
minimax theorem due to Loomis \cite{loomis1946},
which then leads to Gordan's Theorem as an easy additional
step.
As we explain at the end of Section~\ref{s-sti}, it seems
difficult to extend the proof by Loomis to proving LP
duality directly, which was the original aim of this
research.

Section \ref{s-mmlp} presents the classic derivation of LP
duality from the minimax theorem due to Dantzig \cite{dantzig1951}.
Even though its additional assumption looks minor, we show
that it amounts to the Lemma of Tucker \cite{tucker1956},
which, as noted by Tucker \cite[p.~7]{tucker1956}, proves the
Lemma of Farkas.
This shows that the assumption is way too
strong to make Dantzig's derivation useful.

Section~\ref{s-gotu} proves Tucker's Theorem and thus LP
duality from the minimax theorem using Gordan's theorem.
As mentioned, this is distilled from Adler
\cite[section~4]{adler2013}.
In Section~\ref{s-M}, we add another row to Dantzig's game
to obtain a new game where every max-min strategy either
gives a solution to the LP or a certificate that no
optimal solution exists.
Theorems \ref{t-gotu} and~\ref{t-BM} in
Sections~\ref{s-gotu} and~\ref{s-M} are the main results of
this paper.

Section~\ref{s-adler} gives a detailed comparison of our
work with the closely related papers by Adler
\cite{adler2013} and Brooks and Reny \cite{BReny}

In the final Section~\ref{s-minfeas} we present a
little-known gem of a proof of the Lemma of Farkas due to
Conforti, Di Summa, and Zambelli \cite{conforti2007}.
Their theorem states that a system of inequalities $Ax\le b$
is \textit{minimally} infeasible if and only if the
corresponding \textit{equalities} $Ax=b$ are minimally
infeasible.
Because the linear equations are infeasible, a suitable
linear combination of them states $0=-1$, which proves
the Lemma of Farkas in this context.


\section{LP duality and the minimax theorem}
\label{s-state}

Throughout, $m$ and $n$ are positive integers, and
$[n]=\{1,\ldots,n\}$.
All vectors are column vectors.
The $j$th component of a vector $x$ is written $x_j$\,.
All matrices have real entries.
The transpose of a matrix $A$ is written~$A\T$.
Vectors and scalars are treated as matrices of appropriate
dimension, so that a vector $x$ times a scalar~$\alpha$ is
written as $x\alpha$, and a row vector $x\T$ times a scalar
$\alpha$ as $\alpha x\T$.
The matrix $A$ with all entries multiplied by the scalar
$\alpha$ is written as $\alpha A$.
We usually transpose vectors rather than the matrix,
to emphasize that $Ax$ is a linear combination of the
columns of $A$ and $y\T A$ is a linear combination of the rows
of $A$.
The all-zero and the all-one vector are written as
$\0=(0,\ldots,0)\T$ and~$\1=(1,\ldots,1)\T$, their dimension
depending on the context, and the all-zero matrix just
as~0.
Inequalities between vectors or matrices such as $x\ge\0$
hold between all components.

A linear program (LP) in \textit{inequality form} is given by an
$m\times n$ matrix $A$ and vectors $b\in\reals^m$ and
$c\in\reals^n$ and states, with a vector $x\in\reals^n$ of
variables:
\begin{equation}
\label{P}
\maxi_{x} c\T x \quad \subj
Ax\le b,\quad x\ge\0.
\end{equation}
This LP is called \textit{feasible} if there is some
$x\in\reals^n$ that fulfills the constraints $Ax\le b$ and
$x\ge\0$, otherwise \textit{infeasible}.
If there are arbitrarily large values of $c\T x$ with
$Ax\le b$ and $x\ge\0$, then the LP is called
\textit{unbounded}.

With (\ref{P}) considered as the \textit{primal} LP, its
\textit{dual LP} states, with a vector $y\in\reals^m$ of
variables: 
\begin{equation}
\label{D}
\mini_{y} y\T b \quad \subj
y\T A\ge c\T,\quad y\ge \0,
\end{equation}
with feasibility and unboundedness defined accordingly.
An equivalent way of writing the dual constraints in
(\ref{D}) is $A\T y\ge c$, which transposes only the matrix
and can be more readable.

The \textit{weak duality} theorem states that if both primal
and dual LP have feasible solutions $x$ and $y$,
respectively, then their objective function values are mutual
bounds, that is,
\begin{equation}
\label{WD}
c\T x\le y\T b,
\end{equation}
which holds because feasibility implies
$c\T x\le y\T Ax\le y\T b$.
Hence, if there are feasible solutions $x$ and $y$ so that
the two objective functions are \textit{equal}, $c\T x=y\T b$,
then both are optimal.
The (strong) \textit{LP duality} theorem states that this is
always the case if the two LPs are feasible:

\begin{theorem}[LP duality]
\label{t-lp}
If the primal LP $(\ref{P})$ and the dual LP $(\ref{D})$ are
feasible, then there exist feasible $x$ and $y$ with
$c\T x=y\T b$, which are therefore optimal solutions.
\end{theorem} 

A \textit{zero-sum game} is given by an $m\times n$ matrix
$A$ and is played between a \textit{row player}, who chooses
a row $i$ of the matrix, simultaneously with the
\textit{column player}, who chooses a column $j$ of the
matrix, after which the row player receives the matrix entry
$a_{ij}$ from the column player as a \textit{payoff} (which
is a \textit{cost} to the column player).
That is, the row player is the maximizer and the column
player the minimizer.
The rows and columns are called the players' \textit{pure
strategies}.

The players can \textit{randomize} their actions
by choosing them according to a probability distribution,
called a \textit{mixed strategy}.
The other player may know the probability distribution but
not the chosen pure strategy.
The row player is then assumed to maximize his
\textit{expected payoff} and the column player to minimize
her \textit{expected cost}.
We denote the set of mixed strategies of the row player by
\begin{equation}
\label{Y}
Y=\{y\in\reals^m\mid y\ge\0,~\1\T y=1\},
\end{equation}
and of the column player by 
\begin{equation}
\label{X}
X=\{x\in\reals^n\mid x\ge\0,~\1\T x=1\},
\end{equation}
in order to stay close to the LP notation
(normally row and column player are considered as first and
second player, respectively, so that the letters for their mixed
strategies should be in alphabetical order, but this is
already violated with the very common naming of the LP
variables $x\in\reals^n$ and $y\in\reals^m$).

With mixed strategies $y$ and $x$ of row and column player,
the expected payoff to the maximizing row player
and expected cost to the minimizing column player is $y\T Ax$.

The minimizing column player who chooses a mixed strategy
$x$ should expect that the row player responds with a mixed
strategy $y$ (called a \textit{best response}) that
maximizes her payoff $y\T Ax$.
That best-response payoff $\max_{y\in Y}y\T Ax$ is the
weighted sum $\sum_{i\in[m]}y_i (Ax)_i$ of the expected
payoffs $(Ax)_i$ for the rows~$i$ and therefore equal to
their maximum, which in turn is the least upper bound $v$
of these row payoffs. 
That is,
\begin{equation}
\label{Axi}
\max_{y\in Y}~y\T Ax~=~\max_{i\in [m]}~(Ax)_i~=~
\min\,\{v\in\reals\mid Ax\le \1v\,\}\,.
\end{equation}
A \textit{min-max} strategy $x$ of the column player
minimizes this worst-case cost $v$ that he has to pay, that
is, it is an optimal solution to
\begin{equation}
\label{lpv}
\mini_{x,\,v}v\quad \subj A x\le \1v,\quad x\in X
\end{equation}
and then $v$ is called the \textit{min-max value} of the
game.

Similarly, a \textit{max-min} strategy $y$ and the
\textit{max-min value} $u$ is an optimal solution to
\begin{equation}
\label{lpu}
\maxi_{y,\,u}u\quad \subj y\T A\ge u\1\T,\quad y\in Y.
\end{equation}
The minimax theorem of von Neumann \cite{vN1928} states
\begin{equation}
\label{MM}
\max_{y\in Y}~\min_{x\in X}~y\T Ax
~=~v 
~=~
\min_{x\in X} ~\max_{y\in Y}~y\T Ax
\end{equation} 
where the unique real number $v$ is called the
\textit{value} of the game.
Via (\ref{Axi}) and the corresponding expression for
$\min_{x\in X} y\T Ax$ (the best-response cost to
$y\in Y$), we state this as follows.

\begin{theorem}[The minimax theorem]
\label{t-mm}
Consider optimal $x,v$ for $(\ref{lpv})$ and $y,u$ for
$(\ref{lpu})$.
Then $u=v$ (the value of the game), $x$ is a min-max
strategy, and $y$ is a max-min strategy.
\end{theorem} 

The LP (\ref{lpv}) is in \textit{general form} with an
equation $\1\T x=1$ and an unconstrained variable~$v$ (with
$-v$ to be maximized), and so is (\ref{lpu}), which is the
dual LP to (\ref{lpv}) with $u$ as the unconstrained
variable (with $-u$ to be minimized) that corresponds to the
equation for $X$ written as $-\1\T x=-1$.
Since both LPs are feasible, the strong duality theorem
(which also holds for LPs in general form) implies that
their optimal values are equal ($-v=-u$), which
proves Theorem~\ref{t-mm}.

One can avoid stating LPs in general form by ensuring that
the min-max value is positive, by adding a constant $\alpha$
to the payoffs $a_{ij}$\,, which defines a new payoff matrix
$A+\1\alpha\1\T$.
Then for $y\in Y$ and $x\in X$
\begin{equation}
\label{alpha}
y\T (A+\1\alpha\1\T) x
~=~
y\T Ax+ y\T\1\alpha\1\T x 
~=~
y\T Ax+\alpha\,,
\end{equation}
which shows that best responses and min-max and max-min
strategies are unaffected and the corresponding values just
shifted by~$\alpha$.
If all entries of $A$ are positive, then $v>0$ for any
feasible $v$ in (\ref{lpv}).
Division of each row in (\ref{lpv}) by $v$ (where we now
maximize $1/v$) then gives the LP
\begin{equation}
\label{lp1}
\maxi_x \1\T x \quad\subj
Ax\le\1,\quad x\ge\0 
\end{equation}
with its dual
\begin{equation}
\label{lp2}
\mini_y y\T \1 \quad\subj
y\T A\ge\1\T,\quad y\ge\0 \,.
\end{equation}
Both LPs are feasible with nonzero optimal solutions $x$ and
$y$, which give the min-max and max-min strategies 
$xv$ and $yv$ with $v=1/\1\T x=1/\1\T y$ and game value~$v$.

These are the standard ways to derive the minimax theorem
from LP duality 
\cite[section~13-2]{Dantzig}.
Section~\ref{s-mmlp} describes the classical converse
approach, which we show to be incomplete. 

\section{The Lemma of Farkas and LP duality}
\label{s-farkas}

The standard way to prove the LP duality theorem uses the 
Lemma of Farkas \cite{farkas1902}, stated in
(\ref{farkeq}) below, which characterizes
when an inhomogeneous system $Ax=b$ of linear equations has
no solution $x\ge\0$ in nonnegative variables.
Two related theorems are (\ref{fark}) and (\ref{farkine}).
The following proposition asserts how close they are, by
using the respective matrix in different ways (we say
``proves'' rather than ``implies'' because it is not the
same matrix).

\begin{proposition}
\label{p-farkas}
Let $A\in\reals^{m\times n}$ and $b\in\reals^m$.
Then each of the following three assertions proves the
others:
The Lemma of Farkas with equalities and nonnegative variables
\begin{equation}
\label{farkeq}
\not \exists x\in\reals^n~:~Ax=b,~~ x\ge\0
\quad\Leftrightarrow\quad
\exists y\in\reals^m~:~
y\T A\ge\0\T,~~y\T b< 0\,, 
\end{equation}
the Lemma of Farkas with inequalities and nonnegative variables
\begin{equation}
\label{fark}
\not \exists x\in\reals^n~:~Ax\le b,~~ x\ge\0
\quad\Leftrightarrow\quad
\exists y\in\reals^m~:~
y\T A\ge\0\T,~~y\ge\0,~~y\T b< 0\,, 
\end{equation}
and the Lemma of Farkas with inequalities and unconstrained variables
\begin{equation}
\label{farkine}
\not \exists x\in\reals^n~:~Ax\le b
\quad\Leftrightarrow\quad
\exists y\in\reals^m~:~
y\T A=\0\T,~~y\ge\0,~~y\T b< 0\,. 
\end{equation}
\end{proposition}

\myproof
In each of (\ref{farkeq}), (\ref{fark}), (\ref{farkine}) the
direction ``$\Leftarrow$'' is immediate, for example in
(\ref{farkeq}) because 
$y\T A\ge\0\T$ and $Ax=b$, $x\ge\0$ imply $y\T b=y\T Ax\ge0$
which contradicts $y\T b<0$\,.
We therefore only consider ``$\Rightarrow$''.
Condition (\ref{farkeq}) proves (\ref{fark}) by
writing $Ax\le b$ as $Ax+s=b$, $s\ge\0$ for a vector of
\textit{slack variables} $s\in\reals^m$, and then applying
(\ref{farkeq}) to the matrix $[A~~I\,]$ instead of $A$, where
$I$ is the $m\times m$ identity matrix.

Conversely, if there is no solution $x\ge\0$ to $Ax=b$,
that is, to $Ax\le b$ and $-Ax\le-b$, then by
(\ref{fark}) there are
nonnegative $y^+,y^-\in\reals^m$ with
$(y^+)\T A-(y^-)\T A\ge\0\T$ and $(y^+)\T b-(y^-)\T b<0$.
This shows (\ref{farkeq}) with $y=y^+-y^-$.

Condition (\ref{farkine}) follows from
(\ref{fark}) by writing $Ax\le b$ in (\ref{farkine})
as $Ax^+-Ax^-\le b$ with nonnegative $x^+$ and $x^-$.
The converse holds by writing $Ax\le b$, $x\ge\0$ in
(\ref{fark}) as $Ax\le b$, $-x\le\0$ in (\ref{farkine}). 
\endproof

These versions of the Lemma of Farkas are ``theorems of the
alternative'' in that exactly one of two conditions is true,
as in (\ref{farkeq}):
Either there is a solution $x$ to $Ax=b$, $x\ge\0$, or a
solution $y$ to $y\T A\ge\0\T$, $y\T b<0$, but not to both.
We always state such theorems so that ``$\Rightarrow$'' is
the nontrivial direction.

The following is standard (e.g., Gale
\cite[p.~79]{Gale1960}), and similar arguments as used in
the proof will be used repeatedly.

\begin{proposition}
\label{p-fa-lp}
The inequality version $(\ref{fark})$ of the Lemma of Farkas proves LP duality.
\end{proposition}

\myproof
Suppose that the primal LP (\ref{P}) has a feasible solution
$\bar x$ and the dual LP (\ref{D}) has a feasible solution
$\bar y$ and that, contrary to the claim of the LP duality
theorem, there are no feasible $x$ and $y$ so that $c\T
x=y\T b$.
That is, the system of inequalities
\begin{equation}
\arraycolsep.2em
\label{falp}
\begin{array}{rcrcr}
&&Ax&\le&b\\
-A\T y&&&\le&-c\\
b\T y&-&c\T x&\le&0
\end{array} 
\end{equation}
has no solution
$(y,x)\in\reals^m\times\reals^n$ with $y\ge\0$ and $x\ge\0$.
Hence, by (\ref{fark}) (written transposed), there are nonnegative
$(\hat y,\hat x,t)\in\reals^m\times\reals^n\times\reals$
such that 
\begin{equation}
\label{viol}
\arraycolsep.2em
\begin{array}{rcrcrcl}
&-&A\hat x&+&bt&\ge&\0\\
A\T\hat y&&&-&ct&\ge&\0\\
b\T\hat y&-&c\T\hat x&&&<&0\,.
\end{array} 
\end{equation}
If $t>0$ then $\hat x\frac1t$ and $\hat y\frac1t$ are
feasible solutions to the primal (\ref{P}) and dual (\ref{D})
with $\frac1t\hat y\T b<c\T\hat x\frac1t$ in violation of
weak duality (\ref{WD}).
If $t=0$ then $A\hat x\le \0$ and $\hat y\T A\ge\0\T$.
The last inequality in (\ref{viol}) implies that at least
one of the inequalities $\hat y\T b<0$ or $0<c\T\hat x$
holds.
Suppose the latter.
For $\alpha\in\reals$ we have
$A(\bar x+\hat x\alpha)\le b$ and 
$\bar x+\hat x\alpha\ge\0$, but $c\T(\bar x+\hat
x\alpha)\to\infty$ as $\alpha\to\infty$, that is, the
objective function of the primal LP is unbounded,
contradicting its upper bound $\bar y\T b$ from the dual LP.
Similarly, $\hat y\T b<0$ implies that the dual LP is unbounded
and thus the primal LP infeasible, again a contradiction. 
This shows that (\ref{falp}) has a nonnegative solution
$(y,x)$ with $y\T b\le c\T x$ and thus $y\T b=c\T x$ by weak
duality, as claimed.
\endproof

The converse also holds, as well as a useful extension of LP
duality.

\begin{proposition}
\label{p-lp-fa}
The LP duality Theorem~\ref{t-mm} proves $(\ref{fark})$.
Moreover, if the primal LP $(\ref{P})$ is infeasible and the
dual LP $(\ref{D})$ is feasible, then the dual LP is
unbounded.
\end{proposition}

\myproof
Suppose there is no $x\ge\0$ with $Ax\le b$.
Then the LP (with a new scalar variable~$t$)
\begin{equation}
\label{faP}
\maxi_{x,t} -t \quad \subj
Ax-\1t\le b,\quad x\ge\0,~t\ge0
\end{equation}
(which is feasible by choosing $t\ge -b_i$ for all
$i\in[m]$ and $x=\0$) has an optimum solution with $t>0$.
The dual LP to (\ref{faP}) states
\begin{equation}
\label{faD}
\mini_{y} y\T b \quad \subj
y\T A\ge \0\T,~~-y\T\1\ge-1,~~ y\ge \0,
\end{equation}
is feasible with $y=\0$, and therefore has an optimal 
solution $y\ge\0$ with equal objective function value to the
primal, that is, $y\T b=-t<0$.
This shows (\ref{fark}).

To prove the second part, suppose $\bar y\T A\T \ge c\T$ for
some $\bar y\ge\0$.
Then with the preceding $y\ge\0$ such that $y\T b<0$ we have
$(\bar y\T+\alpha y\T)A\ge c\T$ and
$\bar y+y\alpha\ge \0$ and
$(\bar y\T+\alpha y\T)b\to-\infty$ as $\alpha\to\infty$.
\endproof

\section{The theorems of Gordan and Ville}
\label{s-govi}

The Lemma of Farkas with equalities (\ref{farkeq})
characterizes when the inhomogeneous linear equations $Ax=b$ have
no solution $x\ge\0$ in nonnegative variables.
The following Theorem (\ref{gordan}) of Gordan \cite{gordan1873}
for homogeneous equations characterizes when the system $Ax=\0$ has
no nontrivial solution $x\ge\0$.
Its ``inequality version'' (\ref{ville}) is known as the
Theorem of Ville \cite{ville1938}.
Ville's Theorem essentially states the minimax theorem for a
game with positive value.
To prove the minimax theorem from Ville's Theorem, the game
should have its value normalized to zero.
A common way to achieve this is to symmetrize the game \cite{GKT}.
Instead, we shift the payoffs as in (\ref{alpha}) so that
the max-min value is zero.
Note that the min-max and max-min values in (\ref{lpv}) and
(\ref{lpu}) exist without having to assume LP duality.

\begin{proposition}
\label{p-govi}
Let $A\in\reals^{m\times n}$. 
Then the following Theorem $(\ref{gordan})$ of Gordan
proves the Theorem $(\ref{ville})$ of Ville
and vice versa,
and $(\ref{ville})$ proves the minimax
theorem and vice versa:
\begin{eqnarray}
\not \exists x\in\reals^n~:~Ax=\0,~~ x\ge\0,~~ x\ne\0
~~&\Leftrightarrow&~~
\exists y\in\reals^m~:~
y\T A>\0\T\,, 
\label{gordan}
\\
\not \exists x\in\reals^n~:~Ax\le\0,~~ x\ge\0,~~ x\ne\0
~~&\Leftrightarrow&~~
\exists y\in\reals^m~:~
y\T A>\0\T,~~y\ge\0\,. 
\label{ville}
\end{eqnarray}
\end{proposition}

\myproof
Assume (\ref{gordan}) holds. 
We prove (\ref{ville}).
Suppose there is no $x\in\reals^n$ with $Ax\le\0$, $x\ge\0$,
$x\ne\0$. 
Then there is no $x\in\reals^n$ and $s\in\reals^m$ with
$Ax+s=\0$ and $x\ge\0$, $s\ge\0$, and $(x,s)\ne(\0,\0)$
(this clearly holds if $x\ne\0$, and if $x=\0$ then $s=\0$). 
Hence, by (\ref{gordan}), there is some $y\in\reals^m$ with 
$y\T A>\0\T$ and $y>\0$ and thus $y\ge\0$.
This shows the nontrivial direction ``$\Rightarrow$'' in
(\ref{ville}).

Conversely, suppose there is no $x\ge\0$, $x\ne\0$ with
$Ax=\0$ and hence no $x\ge\0$, $x\ne\0$ with $Ax\le\0$ and
$-Ax\le\0$.
Then by (\ref{ville}) there exist $y^+\ge\0$ and $y^-\ge\0$
with $(y^+)\T A+(y^-)\T(-A)>\0\T$,
that is, $(y^+-y^-)\T A>\0\T$, which shows (\ref{gordan})
with $y=y^+-y^-$.

Assume the minimax Theorem~\ref{t-mm} holds for the game
matrix $A$.
The left-hand side of (\ref{ville}) states that the value
$v$ of the game is positive, because otherwise there would
be a mixed strategy $x\in X$ with nonpositive min-max value
$v$ in (\ref{lpv}).
With the optimal $y\in Y$ and $u>0$ in (\ref{lpu}) we have
$y\T A\ge u\1\T>\0\T$ as asserted in (\ref{ville}).

Conversely, assume (\ref{ville}) and consider a game matrix
$A$. 
Let $u$ be its max-min value and $y\in Y$ be a max-min
strategy as in (\ref{lpu}).
Let $A'=A-\1u\1\T$.
Then $y\T A'=y\T A-u\1\T\ge\0\T$.
We claim that $A'x\le\0$ for some $x\in X$.
If not then there is no $x\ge\0$, $x\ne\0$ with $A'x\le\0$
(otherwise scale $x$ so that $x\in X$), and therefore
by (\ref{ville}) we have $y\T A'>\0\T$ for some $y\ge\0$.
Because $y\ne\0$, we can scale $y$ such that $y\in Y$ and
choose $\eps>0$ such that $y\T A'\ge \eps\1\T$ and hence
$y\T A\ge(u+\eps)\1\T$, which contradicts the maximality
of $u$ in (\ref{lpu}).
Hence, there is $x\in X$ with $A'x\le\0$, so 
$A'$ has min-max value zero and therefore $A$ has min-max
value~$u$, which proves the minimax theorem.
\endproof

\section{The theorems of Stiemke and Loomis}
\label{s-sti}

This section is about two proofs of the minimax theorem, for
example in order to use it for proving LP duality.
For historical interest, we first reproduce a short proof of Gordan's
Theorem (\ref{gordan}) by Stiemke \cite{stiemke1915}. 
In modern language, it uses the property that the null space
and row space of a matrix are orthogonal complements, as
stated in (\ref{RS}) below.
We state this property as the following ``theorem of the
alternative'' about the solvability of linear equations
without nonnegativity constraints, which is well known
(e.g., \cite{kuhn1956}).
We also use this lemma in Section~\ref{s-minfeas} for a
short proof of the Lemma of Farkas.

\begin{lemma}
\label{l-alt}
Let $A\in\reals^{m\times n}$ and $b\in\reals^m$. Then
\begin{equation}
\label{linalt}
\not \exists x\in\reals^n~:~Ax=b \quad\Leftrightarrow\quad
\exists y\in\reals^m~:~
y\T A=\0\T,~~y\T b\ne 0\,.
\end{equation}
\end{lemma}

\myproof
We show the nontrivial direction ``$\Rightarrow$''.
Assume that $b$ is not a linear combination of the columns
$A_1,\ldots,A_n$ of $A$.
Let $k$ be the column rank of $A$ and $\{A_j\}_{j\in K}$ be
a basis of the column space of $A$, with $|K|=k\ge0$, and
let $A_K$ be the matrix of these columns.
By assumption, the $m\times (k+1)$ matrix $[A_K~b]$ has rank
$k+1$, which is also its row rank.
Its rows span therefore all of $\reals^{1\times(k+1)}$, in
particular the vector $(\0\T,1)$, that is, $y\T A_K=\0\T$
and $y\T b=1$ for some $y\in\reals^m$.
Any other column $A_j$ of $A$ for $j\not\in K$ is a linear
combination of the basis columns,
$A_j=A_Kz^{(j)}$
for some $z^{(j)}\in\reals^k$, which implies
$y\T A_j=y\T A_Kz^{(j)}=0$.
This shows that overall $y\T A=\0\T$ and $y\T b\ne0$, as
required.
\endproof

\begin{theorem}[Stiemke \cite{stiemke1915}]
\label{t-stiemke}
Let $A\in\reals^{m\times n}$. Then
\begin{equation}
\label{stiemke}
\not \exists y\in\reals^m~:~
y\T A\ge\0\T,~~y\T A\ne\0
\quad\Leftrightarrow\quad
\exists x\in\reals^n~:~Ax=\0\,,~~ x>\0\,. 
\end{equation}
\end{theorem}

\myproof
Define
\begin{equation}
\label{space}
\begin{array}{lcl}
\RS(A)&=&\{y\T A\mid y\in\reals^m\}\,,\\
\NS(A)&=&\{x\in\reals^n\mid Ax=\0\}\,.\\
\end{array} 
\end{equation}
We have for $c\in\reals^n$
\begin{equation}
\label{RS}
c\T\in\RS(A)
\quad\Leftrightarrow\quad
\forall x\in\NS(A)~:~c\T x=0
\end{equation}
because this is equivalent to 
\begin{equation}
\label{RSt}
\exists y~:~y\T A=c\T
\quad\Leftrightarrow\quad
\not\exists x~:~ Ax=\0\,,~~c\T x\ne0\,,
\end{equation}
which (with both sides negated) is the transposed version of
(\ref{linalt}).

The nontrivial direction in (\ref{stiemke}) is
``$\Rightarrow$''.
It states:
Suppose $\0\T$ is the only nonnegative vector in $\RS(A)$.
Then there is some $x\in\NS(A)$ with $x>\0$.
We show this by induction on $n$.
If $n=1$ then the single column of $A$ is $\0$, and we can
choose $x=1$. 
Let $n>1$ and suppose the claim is true for $n-1$.

Case 1.
There is some $a\in\reals^{n-1}$, $a\ge\0$, $a\ne\0$ so that
$(1,-a\T)\in\RS(A)$.
Consider a set of row vectors $(1,-a\T)$,
$(0,a_2\T)$, \ldots, $(0,a_m\T)$ that span $\RS(A)$ (easily
obtained from the rows of~$A$).
There is no $w\in\reals^{m-1}$ such that 
$c\T=\sum_{i=2}^m w_{i-1}\,a_i\T$ is nonnegative and nonzero,
because otherwise $(0,c\T)$ is in $\RS(A)$ and
nonnegative and nonzero.
Hence, by inductive hypothesis, there is some $z\in\reals^{m-1}$,
$z>\0$, such that $a_i\T z=0$ for $2\le i\le m$.
Then $x_1=a\T z>0$, and $x=\twovec{x_1}{z}\in\NS(A)$
by (\ref{RS}) because $(1,-a\T)x=0$ and
$(0,a_i\T)x=0$ for $2\le i\le m$, and $x>\0$.

Case 2.
Otherwise, consider any $y\in\reals^m$ and
let $(c_1,c\T)=y\T A$ with $c\in\reals^{m-1}$.
Then $c\ge\0$ implies $c=\0$, which holds by assumption
if $c_1\ge0$, and if $c_1<0$ and $c\ge\0$, $c\ne\0$ then
$(1,\frac{1}{c_1}c\T)\in\RS(A)$ and Case~1 applies.
By inductive hypothesis, there is some $z\in\reals^{m-1}$,
$z>\0$, such that $A\twovec{0}{z}=\0$.
If $x_1=0$ for all $x\in\NS(A)$ then by (\ref{RS}) we have
$(1,0,\ldots,0)\in\RS(A)$ contrary to assumption.
So there is some $x'\in\NS(A)$ with $x'_1>0$,
and therefore $x=x'\eps+\twovec{0}{z}>\0$ for
sufficiently small $\eps>0$, where $Ax=\0$.
This completes the induction.
\endproof

The preceding theorem is statement I of Stiemke \cite{stiemke1915},
and Gordan's Theorem~(\ref{gordan}) is statement~II.

\begin{proposition}
\label{p-stigo}
Stiemke's Theorem~\ref{t-stiemke} proves Gordan's
Theorem $(\ref{gordan})$. 
\end{proposition}

\myproof
Let $A\in\reals^{m\times n}$.
Let $\{b_1,\ldots,b_k\}$ with $k\ge1$ be a spanning set of $\NS(A)$
and $B=[b_1\cdots b_k]$.
Then for $b$ and $c$ in $\reals^n$ 
\begin{equation}
\label{cNS}
b\in\NS(A)
\quad\Leftrightarrow\quad
b\T \in\RS(B\T)
\end{equation}
and, using (\ref{RS}),
\begin{equation}
\label{cRS}
\begin{array}{ll}
& c\T \in\RS(A)
\\ \Leftrightarrow~~ &
\forall x\in\NS(A)~:~c\T x=0
\\ \Leftrightarrow &
c\T b_i=0\qquad (1\le i\le k)
\\ \Leftrightarrow &
c\T B=\0\T
\\ \Leftrightarrow &
c\in\NS(B\T)\,.
\end{array} 
\end{equation}
Stiemke's Theorem (\ref{stiemke}) applied to $B\T$ instead
of $A$ states
\begin{equation}
\label{sBT}
\not\exists \,b\T\in\RS(B\T)\,,~~b\ge\0\,,~~b\ne\0
\quad\Leftrightarrow\quad
\exists \,c\in\NS(B\T)~:~ c>\0 
\end{equation}
which by (\ref{cNS}) and (\ref{cRS}) is equivalent to
\begin{equation}
\label{G1}
\not\exists \,b\in\NS(A)\,,~~b\ge\0\,,~~b\ne\0
\quad\Leftrightarrow\quad
\exists \,c\T\in\RS(A)~:~ c>\0 
\end{equation}
which is Gordan's Theorem (\ref{gordan}).
\endproof

Via Propositions~\ref{p-govi} and~\ref{p-stigo}, Stiemke's
Theorem~\ref{t-stiemke} therefore proves the minimax
theorem.
Using symmetric games, this was also shown by Gale, Kuhn,
and Tucker \cite{GKT}.

Our favorite proof of the minimax theorem is based on the
following theorem.

\begin{theorem}[Loomis \cite{loomis1946}]
\label{t-loomis}
Let $A$ and $B$ be two $m\times n$ matrices with $B>0$.
Then there exist $x\in X$, $y\in Y$, and $v\in\reals$ such 
that $Ax\le Bxv$ and $y\T A\ge v y\T B$.
\end{theorem}

The case $B=\1\,\1\T$ gives the minimax theorem.
Conversely, the minimax theorem proves Theorem~\ref{t-loomis}
\cite[p.~19]{LRS}:
Because $B>0$, the value of the game $A-\alpha B$ is
negative for sufficiently large~$\alpha$, positive for
sufficiently negative~$\alpha$, is a continuous
function of~$\alpha$, and therefore zero for some $\alpha$,
which then gives Theorem~\ref{t-loomis} with $v=\alpha$.

The following is the proof by Loomis \cite{loomis1946} of
Theorem~\ref{t-loomis} specialized to the minimax theorem.
It is an induction proof about the min-max value $v$ and
max-min value $u$ (which exist, irrespective of LP duality).
It is easy to remember: If the players have optimal
strategies that equalize $v$ and $u$ for all rows and
columns, then $u=v$.
Otherwise (if needed by exchanging the players), there is at
least one row with lower payoff than $v$, which
will \textit{anyhow} not be chosen by the row player.
By omitting this row from the game, the minimax
theorem holds (using a bit of convexity and continuity)
by the inductive hypothesis.

\myproofof{Theorem~\ref{t-mm}}
Consider optimal solutions $v,x$ to $(\ref{lpv})$ and $u,y$
to $(\ref{lpu})$, where
\begin{equation}
\label{weak}
u=u\1\T x\le y\T A x\le y\T \1 v = v.
\end{equation}
We prove $u=v$ by induction on $m+n$. 
It holds trivially for $m+n=2$.
If all inequalities in (\ref{weak}) hold as equalities, then $u=v$.
Hence, assume that at least one inequality is strict, say
$(Ax)_k<v$ for some row $k\in[m]$ (the case for a column is
similar).
Let $\bar A$ be the matrix $A$ with the $k$th row deleted.
By induction hypothesis, $\bar A$ has game value $\bar v$
with $\bar A\bar x\le\1\bar v$ for some $\bar x\in X$,
where it is easy to see that
\begin{equation}
\label{shrink}
\bar v\le v,\qquad\bar v\le u
\end{equation}
because compared to~$A$ the game 
$\bar A$ strengthens the minimizing column player. 

We claim that $\bar v=v$.
Namely, if $\bar v<v$, let $0<\eps\le 1$
and consider the strategy $x(\eps)=x(1-\eps)+\bar x\eps$
where $x\in X$ because $X$ is convex.
Then
\begin{equation}
\label{eps}
\bar A x(\eps)~=~\bar A (x(1-\eps)+\bar x\eps)
\le \1 v(1-\eps)+\1\bar v\eps
=\1(v-\eps(v-\bar v)) 
<\1v\,.
\end{equation}
For the missing row $k$ of $A$ where
$(Ax)_k<v$ we have for sufficiently small~$\eps$
\begin{equation}
\label{Ak}
(Ax(\eps))_k=(Ax)_k(1-\eps)+(A\bar x)_k\eps<v\,.
\end{equation}
Hence, $Ax(\eps)<\1 v$ for some $x(\eps)\in X$,
in contradiction to the minimality of $v$ in (\ref{lpv}).
This shows $v=\bar v$, and, by
(\ref{shrink}), $\bar v\le u\le v=\bar v$
and therefore $u=v$.
This completes the induction.
\endproof 

The proof by Loomis \cite{loomis1946} has been noted (in particular
by von Neumann and Morgenstern
\cite[p.~vi]{vNM}) but is not widely known, and should
be a standard textbook proof (as in \cite[p.~216]{vS22}).
(A better title of Loomis's paper would have been ``An
elementary proof of the minimax theorem'', given that
Theorem~\ref{t-loomis} is not substantially more general.)
It was, in essence, re-discovered by Owen \cite{owen1967}.
However, Owen needlessly manipulates the max-min strategy~$y$;
the proof by Loomis is more transparent. 
Owen's proof is discussed further by Binmore \cite{binmore2004}.

The research in this paper originated with an attempt to
extend the induction proof by Loomis to a direct proof of LP
duality, via the existence of a strictly complementary pair
of optimal strategies in a zero-sum game, applied to
Dantzig's game in (\ref{B}) below.
This existence seems to be difficult to prove within this induction.
For example, the game 
$\left[\begin{matrix}1 & 2 & 0 \\ 1 & 0 & 2
\end{matrix}\right]$
has a max-min and min-max strategy where every pure best
response is played with positive probability (such as both
players mixing uniformly), but also the left column as a
pure min-max strategy.
However, omitting the unplayed second or third column in an
induction would alter the game substantially, because then a
strictly complementary pair has the first column as a unique
min-max strategy, with a positive slack in the column that
was not omitted.

\section{The minimax theorem and LP duality}
\label{s-mmlp}

The following theorem assumes the minimax theorem.

\begin{theorem}[Dantzig \cite{dantzig1951}]
\label{t-51}
Let $A\in\reals^{m\times n}$, $b\in\reals^m$,
$c\in\reals^n$.
Consider the zero-sum game with the payoff matrix $B$
(with $k=m+n+1$ rows and columns) defined by
\begin{equation}
\label{B}
B=\left[\begin{matrix}
0&A&-b\\
-A\T&0&c\\
b\T & -c\T & 0\\
\end{matrix}\right].
\end{equation}
Then $B$ has value zero, with a min-max strategy 
$z=(y,x,t)\in \reals^m \times\reals^n \times\reals$ that is
also a max-min strategy, with $Bz\le\0$.
If $z_k=t>0$ then $x\frac1t$ is an optimal solution to
the primal LP $(\ref{P})$ 
and $y\frac1t$ is an optimal solution to
dual LP $(\ref{D})$.
If $(Bz)_k<0$ then $t=0$ and at least one of the LPs $(\ref{P})$ 
or $(\ref{D})$ is infeasible.
\end{theorem}

\myproof
Because $B=-B\T$, this game is symmetric and its game value
$v$ is zero. 
Let $z=(y,x,t)$.
Then $Bz\le\0$ states $Ax-bt\le\0$, $-A\T y+ct\le\0$, and
$b\T y-c\T x\le0$.
If $t>0$ then $x\frac1t$ and $y\frac1t$ are primal and dual
feasible with $b\T y\frac1t\le c\T x\frac1t$ and therefore
optimal.

If $(Bz)_k<0$, that is, $b\T y-c\T x<0$, then $t>0$ would
violate weak duality, so $t=0$.
Moreover, $Ax\le\0$ and $y\T A\ge\0\T$, and $y\T b<0$ or
$0<c\T x$.
As shown following (\ref{viol}), this implies infeasibility
of at least one of the LPs (\ref{P}) or (\ref{D}).
\endproof

Hence, Theorem~\ref{t-51} seems to show that the minimax
theorem proves LP duality.
The known ``hole'' in this argument is that it is does not
cover the case of a min-max strategy $z$ where $z_k=0$ and
$(Bz)_k=0$, which is therefore uninformative, as noted by
Dantzig \cite[p.~291]{Dantzig}. 
Luce and Raiffa \cite[p.~421]{luceraiffa} claim without
proof
(or forgot a reference, e.g.\ to corollary 3A in
their cited work \citep{goldmantucker1956})
that if $(Bz)_k=0$ for all min-max strategies $z$,
then $\bar z_k>0$ for some max-min strategy $\bar z$.
Because $B$ is skew-symmetric ($B=-B\T$), this would solve
the problem with $\bar z$ as a min-max strategy.
We will show that this assumption is essentially the
Lemma of Tucker \cite[p.~5]{tucker1956} for the case of a
skew-symmetric matrix.
Already for the special case of $B$ in (\ref{B}), this
proves the Lemma of Farkas (\ref{fark})
(see also \cite[theorem~1.1]{broyden2001}), and this defeats
the purpose of proving LP duality from the minimax theorem.

\begin{proposition}
\label{p-Bfa}
Consider $B$ in $(\ref{B})$ with $c=\0$, and suppose that
there is always some $z\ge\0$ with $Bz\le\0$ and
$z_k-(Bz)_k>0$.
Then this proves $(\ref{fark})$.
\end{proposition}

\myproof
Let $z=(y,x,t)$ as described, where $Ax-bt\le\0$ and $-A\T
y\le\0$ and $b\T y\le0$ because $Bz\le\0$, and
$z_k-(Bz)_k=t-b\T y>0$.
Then if $t>0$ we have $Ax\frac1t\le b$, and if $t=0$ 
then $y\T A\ge\0\T$ and $y\T b<0$, which proves
(\ref{fark}). 
\endproof

The Lemma of Tucker comes in several variants.

\begin{proposition}
Let $A\in\reals^{m\times n}$.
Then the following Lemma of Tucker
\label{p-tu}
\begin{equation}
\label{ltuck}
\exists y\in\reals^m, ~x\in\reals^n~:~
y\T A\ge\0\T,\quad
x\ge\0,\quad
Ax=\0,\quad
x_n+(y\T A)_n>0
\end{equation}
proves the following inequality version and vice versa:
\begin{equation}
\label{ituck}
\exists y\in\reals^m, ~x\in\reals^n~:~
y\ge\0,~~
y\T A\ge\0\T,~~
x\ge\0,~~
Ax\le\0,~~
x_n+(y\T A)_n>0\,,
\end{equation}
and similarly its version for a skew-symmetric matrix
$B\in\reals^{k\times k}$, that is, $B=-B\T$: 
\begin{equation}
\label{stuck}
\exists z\in\reals^k~:~
z\ge\0,\quad
Bz\le\0,\quad
z_k-(Bz)_k>0\,. 
\end{equation}
\end{proposition}

\myproof
Applying (\ref{ltuck}) to the matrix $[\,I~~A]$ with the identity
matrix $I$ gives (\ref{ituck}).
For the converse, write $Ax=\0$ as $Ax\le\0$\,, $-Ax\le\0$\,.

Condition (\ref{stuck}) follows from (\ref{ituck}) with
$A=B$ and $z=x+y$ because $-Bz=z\T B$ and $y_n\ge0$ and
$(x\T B)_n\ge0$.
For the converse, use
$B=\left[\begin{matrix}
0&A\\
-A\T&0\\
\end{matrix}\right]$ and $z=\twovec yx$.
\endproof

Tucker \cite[p.~7]{tucker1956} used (\ref{ltuck}) to prove
the Lemma of Farkas in its version (\ref{farkeq}).
Less known, but similarly easy, is that the converse holds
as well.

\begin{proposition}
\label{p-tufa}
The Lemma of Farkas $(\ref{farkeq})$ 
proves Tucker's Lemma $(\ref{ltuck})$.
\end{proposition}

\myproof
%
Let $A=[A_1\cdots A_n]\in\reals^{m\times n}$.
By (\ref{farkeq}),
either $\sum_{j=1}^{n-1}A_jz_j=-A_n$ for some
$z\in\reals^{n-1}$ with $z\ge\0$, in which case let
$x=\twovec{z}{1}$ and $y=\0$, 
or otherwise $y\T A_j\ge0$ for $1\le j<n$ and $y\T(-A_n)<0$ for some
$y\in\reals^m$, in which case let $x=\0$. 
In both cases we have $Ax=\0$ and $x_n+y\T A_n>0$, and
(\ref{ltuck}) holds.
\endproof 


In the next section, we show a proper way of proving LP
duality from the minimax theorem.

\section{Proving Tucker's Theorem from Gordan's Theorem}
\label{s-gotu}

In Tucker's Lemma (\ref{ltuck}), the last
($n$th) column of the matrix $A$ plays a special role, which
can be taken by any other column.
This proves the following stronger version (\ref{tuck})
known as the \textit{Theorem} of Tucker \cite[p.~8]{tucker1956}.

\begin{proposition}
\label{p-tucker}
Let $A\in\reals^{m\times n}$. 
Tucker's Lemma $(\ref{ltuck})$ proves Tucker's Theorem
\begin{equation}
\label{tuck}
\exists y\in\reals^m, ~x\in\reals^n~:~
y\T A\ge\0\T,\quad
x\ge\0,\quad
Ax=\0,\quad
x\T+y\T A>\0\T. 
\end{equation}
\end{proposition}

\myproof
Let $j\in[n]$.
By applying (\ref{ltuck}) to the $j$th column of
$A$ with $j$ instead of $n$, choose $y^{(j)}\in\reals^m$ and
$x^{(j)}\in\reals^n$ such that 
\begin{equation}
\label{jtuck}
(y^{(j)})\T A\ge\0\T,\quad
x^{(j)}\ge\0,\quad
Ax^{(j)}=\0,\quad
x^{(j)}_j+((y^{(j)})\T A)_j>0\,. 
\end{equation}
Then $y=\sum_{j\in[n]}y^{(j)}$ and
$x=\sum_{j\in[n]}x^{(j)}$ fulfill (\ref{tuck}).
\endproof

Tucker's Theorem (\ref{tuck}) is a very versatile
theorem that proves a number of theorems of the alternative 
(see \cite{tucker1956}), for example immediately Gordan's
Theorem $(\ref{gordan})$ or Stiemke's Theorem~\ref{t-stiemke}.

The main Theorem~\ref{t-gotu} of this section shows that
Gordan's Theorem (\ref{gordan}) proves Tucker's Theorem
(\ref{tuck}).
It is based on the following observation.
If $Ax=\0$ and $x\ge\0$, then any $y$ with $y\T A\ge\0\T$
has the property that if $x_j>0$ then $(y\T A)_j=0$ because
otherwise $0=y\T Ax=\sum_{j\in[n]}(y\T A)_jx_j>0$\,.
Hence, (\ref{tuck}) implies that the \textit{support}
\begin{equation}
\label{supp}
S=\supp(x)=\{j\in[n]\mid x_j>0\}
\end{equation}
of $x$ is unique.
The main idea is that the nonnegativity constraints for the
variables $x_j$ for $j\in S$ can be dropped and these
variables therefore be eliminated, which allows applying
Gordan's Theorem to the remaining variables.
The following proof is distilled from the more complicated
computational approach of Adler \cite[section~4]{adler2013}.

\begin{theorem}
\label{t-gotu}
Gordan's Theorem $(\ref{gordan})$ proves Tucker's
Theorem~$(\ref{tuck})$. 
\end{theorem}

\myproof 
Let $A=[A_1\cdots A_n]$.
For any $S\subseteq[n]$ and $J=[n]-S$ write
$A=[A_J~A_S]$ and $x=(x_J,x_S)$ for $x\in\reals^n$.
If $Ax=\0$, $x\ge\0$, $Ax'=\0$, $x'\ge\0$, then
$A(x+x')=\0$, $x+x'\ge\0$, and
$\supp(x+x')=\supp(x)\cup\supp(x')$.
Choose $S$ 
as the inclusion-maximal support
of any $x\ge\0$ such that $Ax=\0$.
Then any $y$ with $y\T A\ge\0\T$ fulfills $y\T A_S=\0\T$
(because otherwise $y\T Ax=y\T A_Sx_S>0$).

On the other hand, (\ref{tuck}) states $x_j+y\T A_j>0$ for
all $j\in[n]$, which requires $y\T A_j>0$ for $j\in
J=[n]-S$.
We now show that there indeed exist $y\in\reals^m$ and
$x=(\0,x_S)$ such that
\begin{equation}
\label{tucker}
y\T A_J>\0\T,~~
y\T A_S=\0\T,~~
Ax=A_Sx_S=\0,~~ x_S>\0\,,
\end{equation}
which implies (\ref{tuck}).
Consider some $\tilde x\ge\0$ with maximum support
$S=\supp(\tilde x)$ such that $A\tilde x=\0$, that is,
$\tilde x_S>\0$.
If $S=[n]$ we are done.
Let $k$ be the rank of~$A_S$.
Suppose $k=m$.
We claim that then $S=[n]$, which implies (\ref{tuck}) with
$y=\0$.
Namely, if $j\in [n]-S$, then $A_j=A_S\hat x_S$ for some
$\hat x_S$ because $A_S$ has full rank,
and therefore $A_j+A_S(\tilde x_S\alpha-\hat x_S)=\0$
where $\tilde x_S\alpha-\hat x_S>\0$ for sufficiently large
$\alpha$, which gives a solution $x\ge\0$ to $Ax=\0$ with
$\supp(x)=\{j\}\cup S$ in contradiction to the maximality of~$S$.

Hence, let $k<m$.
In order to apply Gordan's Theorem~(\ref{gordan}),
we eliminate the variables $x_S$ from the
system $Ax=A_Jx_J+A_Sx_S=\0$ by replacing it with
an equivalent system $CAx=0$ with a suitable invertible
$m\times m$ matrix~$C$.
Let $a_{iS}$ be the $i$th row of $A_S$ for $i\in[m]$.
Suppose for simplicity that the last $k$ rows of $A_S$ are
linearly independent and define the matrix $F$, 
and that for $i=1,\ldots,m-k$ we have $a_{iS}=z^{(i)}F$ for
some row vector $z^{(i)}$ in $\reals^{1\times k}$.
Then the $m\times m$ matrix
\begin{equation}
\label{CC}
C=\left[\begin{matrix}\,
1&\cdots&0&&\kern-2ex-z^{(1)} \\
&\ddots&&&\vdots\\
0&\cdots&1&&\kern-1ex-z^{(m-k)}\kern-1ex\\
0&\cdots&0&1&\cdots&0\\
&\vdots&&&\ddots\\
0&\cdots&0&0&\cdots&1\\
\end{matrix}\,
\right]
\end{equation}
is clearly invertible, and any solution $(x_J,x_S)$ to $A_Jx_J+A_Sx_S=\0$ 
is a solution to
\begin{equation}
\label{CA}
CA_Jx_J+CA_Sx_S=\0
\end{equation}
and vice versa, with
\begin{equation}
\label{DEF}
CA_J=\left[\begin{matrix}D\\E\end{matrix}\right], \quad 
CA_S=\left[\begin{matrix}0\\F\end{matrix}\right]
\end{equation}
where $D\in\reals^{(m-k)\times|J|}$,
$E\in\reals^{k\times|J|}$,
and
$F\in\reals^{k\times|S|}$.

Suppose there is some $x_J\in\reals^{|J|}$ with
\begin{equation}
\label{Go}
Dx_J=\0\,,\quad
x_J\ge\0,~~
x_J\ne\0\,.
\end{equation}
Because $F$ has rank~$k$ there exists $x_S$ so that 
$Fx_S=-Ex_J$.
Then $Ex_J+Fx_S=\0$ and
hence $CA_Jx_J+CA_Sx_S=\0$
and thus $A_Jx_J+A_Sx_S=\0$.
With
$x(\alpha)=(x_J,x_S+\tilde x_S\alpha)$ we have
$Ax(\alpha)=\0$ (because $A_S\tilde x_S=\0$)
and $x(\alpha)\ge\0$ for $\alpha\to\infty$, where
$x(\alpha)$ has larger support that $S$, but $S$ was
maximal.
Hence, there is no $x_J$ so that (\ref{Go}) holds.
By Gordan's Theorem~(\ref{gordan}), there is some
$w\in\reals^{m-k}$ with $w\T D>\0\T$, that is,
\[
(w\T,\0\T)\left[\begin{matrix}D\\E\end{matrix}\right]>\0\T, \quad 
(w\T,\0\T)\left[\begin{matrix}0\\F\end{matrix}\right]=\0\T.
\]
With $y\T=(w\T,\0\T)C$ and (\ref{DEF}), this implies
(\ref{tucker}) with $x=\tilde x$, as claimed.
\endproof

Because the minimax theorem proves Gordan's Theorem (see
Proposition~\ref{p-govi}), it proves Tucker's
Theorem~(\ref{tuck}) and Tucker's Lemma (\ref{ltuck}) and
the Lemma of Farkas and therefore LP duality.

Instead of the minimax theorem we can by
Proposition~\ref{p-stigo} use Stiemke's
Theorem~\ref{t-stiemke} to prove Gordan's
Theorem~(\ref{gordan}).
The short proof by Tucker \cite[p.~5--7]{tucker1956} of his Lemma
(\ref{ltuck}) has some structural similarities to Stiemke's
proof but uses more explicit computations.

We conclude this section to show how Tucker's Theorem
proves, as one of its main applications
\cite[theorem~6]{tucker1956}, the condition of
\textit{strict complementarity} in linear programming.
For the LP (\ref{P}) and its dual LP
(\ref{D}), a feasible pair $x,y$ of solutions is optimal if
and only if we have equality in (\ref{WD}), that is,
$c\T x=y\T Ax= y\T b$, which means
\begin{equation}
\label{cs}
y\T(b-Ax)=0\,,\qquad(y\T A-c\T)x=0\,.
\end{equation}
This orthogonality of the nonnegative vectors $y$ and
$b-Ax$, and of $y\T A-c\T$ and~$x$, means that they are
complementary in the sense that in each component
\textit{at least one} of them is zero:
\begin{equation}
\label{csij}
y_i(b-Ax)_i=0
\quad(i\in[m]),
\qquad
(y\T A-c\T)_j\,x_j=0
\quad(j\in[n]),
\end{equation}
also called ``complementary slackness''.
The following theorem asserts \textit{strict}
complementarity, namely that if (\ref{P}) and (\ref{D}) are
feasible, then they have feasible solutions $x$ and $y$
where \textit{exactly one} of each component in (\ref{csij})
is zero.

\begin{proposition}
\label{p-strict}
If the LPs $(\ref{P})$ and $(\ref{D})$ are feasible, then
they have optimal solutions $x$ and $y$ such that
$(\ref{cs})$ holds and
\begin{equation}
\label{strict}
y+(b-Ax)>\0,\qquad x\T+(y\T A -c\T)>\0\T.
\end{equation}
\end{proposition}

\myproof
Optimality of $x$ and $y$ means $c\T x=y\T b$ and therefore
(\ref{cs}).
Similar to Proposition~\ref{p-tu} and (\ref{stuck}),
Tucker's Theorem (\ref{tuck}) proves that for a
skew-symmetric matrix $B$ there is some $z$ such that 
\begin{equation}
\label{sT}
z\ge\0\,,\quad
Bz\le\0\,,\quad
z-Bz>\0\,. 
\end{equation}
Applied to the game matrix $B$ in (\ref{B}), because the LPs
are feasible, this gives a solution $z=(y',x',t')$ with
$t'>0$, where $y=y'\frac1t$ and $x=x'\frac1t$ fulfill
(\ref{strict}).
\endproof

The proof of Proposition~\ref{p-strict} demonstrates a very
good use of Dantzig's game $B$ in (\ref{B}).
Geometrically, the LP solutions $x$ and $y$ are then in the
relative interior of the set of optimal solutions.
Unless this set is a singleton, $x$ and $y$ are not unique,
but their supports $\supp(x)$ and $\supp(y)$ are unique,
shown similarly to the initial argument in the proof of
Theorem~\ref{t-gotu}.

\section{Extending Dantzig's game}
\label{s-M}

In this section, we give a longer but more constructive
proof of LP duality from the minimax theorem.
We present a natural extension of
Dantzig's game $B$ in (\ref{B}) by adding an extra row
to~$B$, giving the game $B_M$ in (\ref{BM}) below.
The aim is to ``enforce'' the last column of $B$ to be
played with positive probability~$t$ if that is possible.
Any max-min strategy for $B_M$ gives not only information
about solutions to the LPs (\ref{P}) and (\ref{D}) if both
are feasible, but also a certificate in (\ref{witn}) if not.

\begin{theorem}
\label{t-BM}
There is some $M\in \reals$ with the following properties:
If both the primal LP $(\ref{P})$ and its dual $(\ref{D})$
are feasible, then  they also have respective feasible
solutions $x$ and $y$ with $\1\T x+\1\T y+1\le M$.
Moreover, consider the zero-sum game
\begin{equation}
\label{BM}
B_M=\left[\begin{matrix}
0&A&-b\\
-A\T&0&c\\
b\T & -c\T & 0\\
\1\T & \1\T & -M \\
\end{matrix}\right] 
\end{equation}
with value $v$. Then $v\ge0$, and

\rmitem{(a)}
$v=0$ with min-max strategy $(y,x,t)$ and
max-min strategy $(y,x,t,0)$ for $B_M$ if and only if $(\ref{P})$ and
$(\ref{D})$ are feasible, in which case $x\frac1t$ is
optimal for $(\ref{P})$ and $y\frac1t$ is optimal for
$(\ref{D})$.

\rmitem{(b)}
If $v>0$ with max-min strategy $(y,x,r,s)$ for $B_M$,
then $r=0$, $s=v$, and 
\begin{equation}
\label{witn}
A x\le\0,\quad
x\ge\0, \quad
A\T y\ge\0,\quad
y\ge\0, \quad
b\T y-c\T x<0\,, 
\end{equation}
which proves that 
$(\ref{P})$ or $(\ref{D})$
is infeasible.
Moreover, $v<1$, and the smallest number $w$ such that 
\begin{equation}
\label{Aw}
A\bar x\le b+\1 w,
\quad
\bar x\ge\0, \quad
-A\T \bar y\le-c+\1 w,
\quad \bar y\ge\0
\end{equation}
has feasible solutions $\bar x$ and $\bar y$ is given by
\begin{equation}
\label{wv}
w=\frac{M+1}{1/v-1}\,.
\end{equation}
\rmitem{(c)}
If the entries of $A,b,c$ are rational numbers, let $\alpha$
be the maximum of the absolute value of the numerators of
these numbers, let $\beta$ be the maximum
denominator, and $\ell=m+n+1$.
Then a suitable choice of $M$ is 
\begin{equation}
\label{Mell}
M=\ell!\,\ell\alpha^\ell\beta^{\ell^2+\ell}+1,
\end{equation}
which in bit-size is polynomial in the bit-size of $A,b,c$.
\end{theorem}

We first discuss Theorem~\ref{t-BM}.
We will prove it (in Theorem~\ref{t-DM} below) without using
LP duality, which will therefore be an alternative proof of
LP duality from the minimax theorem.
Although this proof is longer than that of
Theorem~\ref{t-gotu}, it provides a reduction of the problem
of solving an LP (in the sense of providing an optimal
solution or a certificate that the LP is unbounded or
infeasible) to the problem of solving a zero-sum game.
This reduction is new, as discussed further in
Section~\ref{s-adler}.

Some observations in Theorem~\ref{t-BM} are immediate:
The value $v$ of $B_M$ is nonnegative because the row player
can ignore the last row and play as in Dantzig's game~$B$ in
(\ref{B}).
Furthermore, if $v=0$, then the second-to-last
row in $B_M$ states $\1\T y+\1\T x -Mt\le 0$ for any min-max
strategy $(y,x,t)$, which means $t>0$.
That strategy can be used as a max-min strategy (with the
last row of $B_M$ unplayed), with optimal solutions
$x\frac1t$ and $y\frac1t$ to (\ref{P}) and (\ref{D}).
For the converse, however, (\ref{P}) and (\ref{D}) may have
feasible solutions $x$ and $y$, respectively, but none of
them fulfill $c\T x\ge y\T b$ unless we assume the LP
duality theorem (which then proves~(a)).
In order to avoid using strong LP duality, we have to argue
more carefully, as done in Theorem~\ref{t-DM} below.
Also, the optimal strategies $x\frac1t$ and $y\frac1t$ fulfill
$\1\T y\frac1t+\1\T x\frac1t\le M$, so this constraint
does not (and must not) affect feasibility of (\ref{P}) and
(\ref{D}).

Theorem~\ref{t-BM}(b) gives a certificate that at least one
of the LPs (\ref{P}) and (\ref{D}) is infeasible, if that is
the case, via any max-min strategy $(y,x,r,s)$.
Then (\ref{witn}) holds (which follows from $r=0$ and
$s=v$), which implies $c\T x>0$ or $b\T y<0$ (or both) and
thus unbounded solutions to (\ref{P}) or (\ref{D}),
respectively, if either LP is feasible (and then the other
LP is not).
Furthermore, the value $v$ of $B_M$ defines, in a strictly
monotonic relation (\ref{wv}), the minimal constant $w$ in
(\ref{Aw}) added as extra slack to the right-hand sides that
makes both LPs feasible.
Given $A,b,c$, the value of $w$ in (\ref{Aw}) is clearly
unique (and finite and independent of~$M$),
whereas the game value $v$ of $B_M$ depends on~$M$.


Theorem~\ref{t-BM}(c) shows that a suitable constant $M$ can
be found by identifying the largest numerator (in absolute
value) $\alpha$ and denominator $\beta$ of the entries in
$A,b,c$ if these are given as rational numbers.
(A similar a priori bound is known if these entries are
algebraic numbers \cite[p.~172]{adler2013}, but not if they
are general real numbers.)
Although $M$ in (\ref{Mell}) is large, its description as a
binary number is of polynomial size in the description of
$A,b,c$.
The conversion of the LPs (\ref{P}) and (\ref{D}) to the
game matrix $B_M$ is therefore a polynomial ``Karp-type''
reduction, where any minimax solution of $B_M$ either solves
the LPs or proves the infeasibility of at least one of them.

Finding $M$ as in Theorem~\ref{t-BM} uses the following
well-known concept.
A \textit{basic} solution $x$ to $Ax=b$ is given by a
solution $x$ where the columns $A_j$ of $A$
with $x_j\ne0$ are linearly independent, which then
determine uniquely the solution~$x$.
These columns are then easily extended to a basis of the
column space of $A$ and define a \textit{basis matrix}.
If $A$ has full row rank~$m$, then a basis has size~$m$, and
the basis matrix is an invertible $m\times m$ matrix.
A basic \textit{feasible} solution $x$ also fulfills
$x\ge\0$.
A basic feasible solution to inequalities $Ax\le b$ (and
$x\ge\0$) is meant to be a basic feasible solution to the
system $Ax+p=b$ (and $x,p\ge\0$), which has full row rank.

Part (b) in the following lemma and its proof are due to Ilan
Adler (personal communication, 2022).

\begin{lemma}
\label{l-cara}
Let $A\in\reals^{m\times n}$,
$b\in\reals^{m}$,
$c\in\reals^{n}$.

\rmitem{(a)} 
If $Ax=b$, $x\ge\0$ has a feasible solution $x$,
then it also has a basic feasible solution.

\noindent
Furthermore, suppose the LP: minimize $c\T x$ subject to $Ax=b$, $x\ge\0$
is feasible and has 

\vskip-\parskip
\noindent
a known lower bound $\lambda$, that is, $c\T x\ge
\lambda$ for all feasible~$x$.
Then
\rmitem{(b)} 
for every feasible solution $x$ to $Ax=b$, $x\ge\0$ there is
a basic feasible solution $x^*$ with $c\T x^*\le c\T x$, 
\rmitem{(c)} 
and ~$\min\{c\T x^*\mid Ax^*=b,~x^*\ge\0,~x^*$ is basic$\} =
\min\{c\T x\mid Ax=b,~x\ge\0\}$.
\end{lemma}

\myproof
Choose a feasible $x$ to $Ax=b$, $x\ge\0$ with minimal support.
Then the columns $A_j$ of
$A$ for $x_j>0$ are linearly independent:
Namely, if $Az=\0$ for some $z\ne\0$
where $z_j\ne0$ implies $x_j>0$, let $P=\{j\mid z_j>0\}$
where $P\ne\emptyset$ (otherwise replace $z$ by $-z$).
Then with
\begin{equation}
\label{cara}
\alpha =\min\{{x_j}/{z_j}\mid j\in P\},\qquad x'=x-z\alpha
\end{equation}
we have $Ax'=b$, $x'\ge\0$, and $x'$ of
smaller support than~$x$.
Hence, no such $z$ exists, which proves the claimed linear
independence.
This shows (a).

To show (b), suppose $Ax=b$ and $x\ge\0$ and $x$ is not
basic, with $Az=\0$ for some $z\ne\0$ where $z_j\ne0$
implies $x_j>0$ as before.
If $c\T z<0$, or if $c\T z=0$ and $z\le\0$, replace
$z$ by $-z$.
Let $P=\{j\mid z_j>0\}$.
Then $P\ne\emptyset$,
which holds if $c\T z=0$ because $z\ne\0$, and
if $P=\emptyset$ and $c\T z>0$ then $z\le\0$, and $x-z\alpha$
is feasible but $c\T(x-z\alpha)$ is arbitrarily negative as
$\alpha\to\infty$, which contradicts boundedness.
Then with $\alpha$ and $x'$ as in (\ref{cara}), $x'$ has
smaller support than $x$ and $c\T x'\le c\T x$.
If $A$ has $n$ columns, then this process terminates after
at most $n$ steps with a basic feasible solution $x^*$
with $c\T x^*\le c\T x$, as claimed.

Part (c) follows from (b) because there are finitely many
basic feasible solutions, so the minimum on the left exists,
and the minimum on the right also exists and equals its
infimum. 
\endproof

With the added equation $\1\T x=1$, 
Lemma~\ref{l-cara}(a) is \textit{Carath\'eodory's theorem}:
Any convex combination $b$ of points in $\reals^m$ is already
the convex combination of a suitable set of at most $m+1$ of
these points \cite[p.~200]{caratheodory1911}. 


We prove Theorem~\ref{t-BM} using the following Theorem~\ref{t-DM}
(mostly to simplify notation) applied to 
\begin{equation}
\label{Cd}
C=\left[\begin{matrix}
0&A\\
-A\T & 0\\
\end{matrix}\right],
\qquad
d=\left[\begin{matrix}
~b\\
-c\\
\end{matrix}\right].
\end{equation}
The proof of Theorem~\ref{t-DM} does \textit{not} use strong LP
duality.

\begin{theorem}
\label{t-DM}
Let $C\in\reals^{k\times k}$ such that $C=-C\T$, and 
$d\in\reals^k$.
Let $(z,w)=(z^*,w^*)\in\reals^k\times\reals$ be a basic
feasible solution that minimizes $w$ subject to
\begin{equation}
\label{basic}
Cz-\1w\le d,\quad
d\T z-w\le 0,\quad
z\ge\0,\quad w\ge0,
\end{equation}
and let $M\in\reals$ with 
\begin{equation}
\label{M}
\1\T z^*+1\le M
\,.
\end{equation}
Consider the zero-sum game
\begin{equation}
\label{DM}
D_M=\left[\begin{matrix}
C&-d\\
\,d\T & 0\\
\1\T & -M \\
\end{matrix}\right] 
\end{equation}
with game value~$v$.
Then $v\ge0$ and
\rmitem{(a)}
$v=0$ if and only if $w^*=0$.
If $w^*=0$, let 
$t=\frac1{\1\T z^*+1}$ and $z=z^*t$.
Then $(z,t)$ is a min-max strategy and $(z,t,0)$ is a
max-min strategy for $D_M$.

\rmitem{(b)}
Suppose $v>0$.
Then every max-min strategy $(q,r,s)$ of $D_M$ fulfills 
$r=0$, $s=v$,
and
\begin{equation}
\label{witness}
Cq\le\0,\quad 
d\T q<0,
\end{equation}
which proves that there is no $z\ge\0$ with $Cz\le d$.
\end{theorem}

\myproof 
In the following, letters (and their decorated versions)
$q$ and $z$ denote vectors in $\reals^k$, and
$r,s,t,u,v,w$ denote scalars in $\reals$.

The system (\ref{basic}) is feasible, for example with
$z=\0$ and large enough~$w$, and $w$ is bounded from below,
so that (\ref{basic}) has an optimal basic feasible solution
$(z^*,w^*)$ by Lemma~\ref{l-cara}(c).

We have $v\ge0$, because the game matrix
\begin{equation}
\label{D=}
D=\left[\begin{matrix}
C&-d\\
\,d\T & 0\\
\end{matrix}\right] 
\end{equation}
is skew-symmetric and has game value $0$, so by adopting any
max-min strategy for $D$ and not playing the last row in
$D_M$ the row player will get at least~$0$.

For the ``if'' part of case (a), if $w^*=0$ then
with $t=\frac1{\1\T z^*+1}$ and $z=z^*t$
we have $\1\T z-Mt\le -t<0$ by (\ref{M}).
This shows that $(z,t)$ is a min-max strategy 
and $(z,t,0)$ a max-min strategy for $D_M$, and $v=0$.
For the ``only if'' part, if $v=0$ then a min-max
strategy $(z',t)$ for $D_M$ requires $t>0$ to get a
nonpositive cost in the last row, and then $z=z'\frac1t$
solves (\ref{basic}) with $w=0$.

To show (b), let $v>0$.
The following properties hold for any optimal
strategies of $D_M$.
The min-max value of $D_M$ with min-max strategy $(z,t)$ is
the smallest real number $v$ such that
\begin{equation}
\label{P'}
\arraycolsep.2em
\begin{array}{rcrcrl}
Cz&-&dt&\le&\1v\\
d\T z&&&\le&v\\
\1\T z&-&Mt&\le&v\\
\1\T z&+&t&=&1\\
z&,&t&\ge&\0&. 
\end{array} 
\end{equation}
The max-min value of $D_M$ with max-min strategy
$(q,r,s)$ is the largest $v$ such that
\begin{equation}
\label{D'}
\arraycolsep.2em
\begin{array}{lclclcl}
q\T C&+&rd\T &+&s\1\T&\ge&v\1\T\\
q\T(-d)&&&-&sM&\ge&v\\
q\T\1&+&r&+&s&=&1\\
q&,&r&,&s&\ge&\0\,. 
\end{array} 
\end{equation}

Then $0<s<1$
because if $s=0$ then $(q,r,0)$ would be a max-min strategy
for the symmetric game $D$ in (\ref{D=}) with max-min value
$v>0$ which is not possible, and if $s=1$ then the last row
of $D_M$ alone would be a max-min strategy for $D_M$, but
that row has the negative entry $-M$.

Because $s>0$, we have $\1\T z-Mt=v$ in (\ref{P'}), and,
using $\1\T z=1-t$,
\begin{equation}
\label{t}
v=1-(M+1)t\,.
\end{equation}
We show that $v\le s$.
If $v>s$, then by (\ref{D'}), 
\begin{equation}
\label{DD}
\arraycolsep.2em
\begin{array}{lclcll}
q\T C&+&rd\T&\ge&(v-s)\1\T\\
q\T(-d)&&&\ge&v+Ms\\
\end{array} 
\end{equation}
which would define a max-min strategy
$(q\frac 1{1-s},\frac r{1-s})$ with positive
max-min value for the symmetric game~$D$, a contradiction.

Hence, $0<v\le s<1$ and by (\ref{t}), 
\begin{equation}
\label{tv}
t=\frac{1-v}{M+1}>0\,. 
\end{equation}
Then (\ref{P'}) implies
\begin{equation}
\label{Czt}
\textstyle
Cz\frac1t\le d+\1\frac vt\,,
\quad
d\T z\frac1t\le \frac vt\,,
\end{equation}
and therefore
\begin{equation}
\label{w*}
w^*\le\frac vt = \frac{v(M+1)}{1-v}
\,.
\end{equation} 

In order to show that every max-min strategy $(q,r,s)$ for
$D_M$ is of the form $(q,0,v)$, we will in essence use weak
duality.
We write $s=u+v$ with $u\ge0$
(we know $s\ge v$) and let $v$ in (\ref{D'}) be
\textit{fixed} where we now in essence maximize~$u$.
That is, we consider the constraints
\begin{equation}
\label{Dbar}
\arraycolsep.2em
\begin{array}{lclclcl}
q\T C&+&rd\T&+&u\1\T&\ge&\0\T\\
q\T(-d)&&&-&uM&\ge&v(M+1)\\
q\T\1&+&r&+&u&=&1-v\\
q&,&r&,&u&\ge&\0~. 
\end{array} 
\end{equation}
They have solutions with the current max-min strategy
$(q,r,s)$ and $u=s-v$.
We use that $Cz^*-d-\1w^*\le\0$ and $d\T z^*-w^*\le 0$ in
(\ref{basic}), and $-1\ge \1\T z^*-M-w^*$ by (\ref{M}),
and $v(M+1)-(1-v)w^*\ge0$ by (\ref{w*}) 
in the following chain of inequalities, obtained by multiplying 
the first inequality in (\ref{Dbar}) by $z^*$, the second
by~1, and the equation by $-w^*$ and summing up:
\begin{equation}
\label{chain}
\begin{array}{rcl}
0&\ge& -u\\
&\ge& q\T(Cz^*-d-\1w^*)+r(d\T z^*-w^*)+u(\1\T z^*-M-w^*)\\
&\ge& v(M+1)-(1-v)w^*\ge0\,.\\ 
\end{array}
\end{equation}
Hence, all inequalities hold as equalities, 
in particular 
\begin{equation}
\label{w*=}
w^*
= \frac{M+1}{1/v-1}
\end{equation} 
and $u=0$.
This shows $s=v$ in any solution $(q,r,s)$ to (\ref{D'}).
In addition,
$q\T C + rd\T\ge \0\T$, that is,
$Cq-dr\le\0$, and $q\T d\le-v(M+1)<0$.
The skew-symmetry of $C$ implies $q\T C q=(q\T C q)\T=q\T
C\T q=-q\T C q$ and therefore $q\T Cq=0$, for any~$q$.
If we had $r>0$ then $Cq\frac1r\le d$
and $0 = q\T C q\frac1r\le q\T d < 0$,
a~contradiction, which shows $r=0$.
This shows $Cq\le\0$ and $d\T q<0$ as claimed in (\ref{witness}).
In turn, this shows that there is no $z\ge\0$ with $Cz\le d$,
because this would imply
$
0\le z\T(-Cq)=z\T C\T q = q\T C z\le q\T d < 0\,.
$
\endproof

\myproofof{Theorem~\ref{t-BM}}
We apply Theorem~\ref{t-DM} to $C$ and $d$ in (\ref{Cd}).
Let $v$ be the value of the game $D_M$.
Then by Theorem~\ref{t-DM}(a), $v=0$ implies feasibility and
optimality of the LPs (\ref{P}) and (\ref{D}).
Conversely, suppose that (\ref{P}) and (\ref{D}) are
feasible.
Then $v=0$, because if $v>0$ then
(\ref{witness}) contradicts feasibility.
This shows part (a) in Theorem~\ref{t-BM},
and also part (b) via (\ref{w*=}). 

To show Theorem~\ref{t-BM}(c), suppose first that $\beta=1$,
that is, all entries of $A,b,c$ are integers.
The system (\ref{basic}) has $\ell$ rows, and written as
equations with slack variables has entries from $A,b,c$ or
$0,1,-1$.
Any basic solution is uniquely determined by the basis
matrix, where each variable is the quotient of two
determinants where the denominator is at least~1 and the
numerator bounded in absolute value by $\ell!\,\alpha^\ell$.
Only the $\ell$ basic variables can be nonzero, so that we
can choose $M=\ell!\,\ell\alpha^\ell+1$ by (\ref{M}).
See also \cite[p.~30]{papasteig} or \cite[p.~172]{adler2013}; 
I did not find the next description, clearly standard,
if $\beta>1$.

If $\beta>1$, multiply each column of $\twovec C{d\T}$ and
$\twovec d0$ in (\ref{basic}) with the least common multiple
of the denominators
in that column, called the \textit{scale factor} $\sigma_j$
for that column $j$ (with $j=0$ if the column is~$d$).
This gives an integral system where each basic solution 
has to be changed by multiplying each variable in column~$j$
with its scale factor $\sigma_j$ and dividing it by
$\sigma_0$ to give the solution to the original system.
Each entry of the integral system has been multiplied by at most
$\beta^\ell$ (this is an overestimate because each column of
$C$ in (\ref{Cd}) has $m$ or $n$ zeros), so we have to
replace $\alpha^\ell$ by $\alpha^\ell(\beta^{\ell})^\ell$,
with the extra factor $\beta^\ell$ for the re-scaling of the
variables, which shows (\ref{Mell}).
The number of bits to represent $M$ is its binary logarithm,
which is polynomial in $\ell$ and in the bit-sizes of $\alpha$
and~$\beta$, and hence in the bit-size of $A,b,c$.
\endproof

\section{Discussion and related work}
\label{s-adler}

Because Dantzig's proof in Theorem~\ref{t-51} works for
generic LPs, a first question is if genericity can be
achieved by perturbing a given LP.
However, this may alter its feasibility.
For example, consider the LP of
maximizing $x_2$ subject to $x_2\le1$, $x\ge\0$,  $x\in\reals^2$.
The corresponding game $B$ in (\ref{B}) has an all-zero row
and column, which when played as an optimal pure-strategy
pair does not play the last column ($t=0$).
The LP has optimal solutions $(x_1,1)$ for any $x_1\ge0$.
However, maximizing the perturbed objective function $\eps
x_1+x_2$ (for some small $\eps>0$) with the same constraints
gives an unbounded LP.
Hence, there is no obvious way of perturbing the LP to make
Dantzig's proof generally applicable.

The closest related works to ours are Adler \cite{adler2013}
and Brooks and Reny \cite{BReny}.
We continue here our discussion from the introduction.

A main goal of Adler \cite{adler2013} is to reduce the
computational problem of solving an LP (in the sense of
finding an optimal solution or proving there is none) to the
problem of solving a zero-sum game by means of a strongly
polynomial-time reduction.
Adler considers the feasibility problem with equalities,
that is, to find $x\in\reals^n$ such that
\begin{equation}
\label{feaseq}
Ax=b,\qquad x\ge\0\,,
\end{equation}
for an $m\times n$ matrix $A$, or to show that no such $x$
exists.
He constructs a symmetric game with $m+n+3$ rows and
columns.
An optimal strategy to that game produces either a solution
to (\ref{feaseq}), or a vector $y\in\reals^m$ such that
$y\T A\ge\0\T$ and $y\T b<0$ (which by (\ref{farkeq}) shows that
(\ref{feaseq}) is infeasible), or some $\tilde x\ne\0$ such
that $A\tilde x=\0$ and $\tilde x\ge\0$.
The first two cases answer whether (\ref{feaseq}) is
feasible or not.
In the third case, $Ax=b$ is replaced by an equivalent
system where the variables $x_{S}$ in the support $S$
(written $J^+$ in \cite{adler2013})
of~$\tilde x$ are eliminated.
In a solution to that equivalent system, the variables
$x_{S}$ can be substituted back, and irrespective of their
sign can be replaced by $x_{S}+\tilde x_{S}\alpha$ for
sufficiently large $\alpha$ to find a solution to
(\ref{feaseq}). 
(The latter step is implicit in the claim (10b) of
\cite[p.~173]{adler2013} and attributed to \cite{Dantzig}
but without a page number; I could not find it and found
these computations the hardest to follow.)
Repeating this at most $n$ times, with corresponding calls
to solving a zero-sum game, then answers the feasibility
problem.
This is known as a ``Cook-type'' reduction.
It also leads to a proof of Tucker's Theorem from Gordan's
Theorem in \cite[section~4]{adler2013}, which we have given
in a more direct way in Theorem~\ref{t-gotu}.

A different ``Karp-type'' reduction uses only a single step
from the feasibility problem (\ref{feaseq}) to solving a
zero-sum game, by adding a constraint $\1\T x\le M$
where $M$ is large enough to not affect feasibility.
If the entries of $A$ and $b$ are algebraic numbers (in
particular, integers), they determine an explicit bound on
$M$ of polynomial encoding size \cite[p.~172]{adler2013}.

We have done the same in Theorem~\ref{t-BM} above.
However, our game $B_M$ is directly derived from the
original LPs (\ref{P}) and (\ref{D}) defined by inequalities
(also first considered by Adler) with a single extra row
added to Dantzig's original game~$B$ in (\ref{B}), rather
than converting them to equalities as in (\ref{feaseq})
(with a new, larger matrix~$A$) and then back to
inequalities to construct an even larger symmetric game.
As an additional, new property, Theorem~\ref{t-BM}(b) shows
that a max-min strategy of~$B_M$ provides a certificate that
the LPs are infeasible if that is the case.


Brooks and Reny \cite{BReny} prove the following theorem.
For any matrix $D$, let $\|D\|$ be the maximum absolute
value of its entries.

\begin{theorem}[Brooks and Reny \cite{BReny}]
\label{t-BR}
Consider the LPs
$(\ref{P})$ and $(\ref{D})$.
Let $r$ be the rank of the matrix
\begin{equation}
\label{hatA}
\hat A=
\left[\begin{matrix}
0&-A\T \\
A&0   \\
-c\T &b\T 
\end{matrix}~\right] 
\end{equation}
and let
\begin{equation}
\label{dalpha}
\alpha=2r^2\max\{\|b\|,\|c\|\}\max_W\|W^{-1}\|+1, 
\end{equation}
where the second maximum is taken over all invertible
sub-matrices $W$ of $\hat A$.
Then 
for the game $P$ with $n+m+1$ rows and columns
\begin{equation}
\label{PP}
P=
\left[\begin{matrix}
0&-\alpha A\T & \0 \\
\alpha A&0  & \0 \\
-\alpha c\T & \alpha b\T & 0
\end{matrix}~\right]
+
\left[\begin{matrix}
c\\-b\\0
\end{matrix}\right]
\1\T 
\end{equation}
either
\rmitem{(a)}
the value of $P$ is zero, and then
for a min-max strategy $(x^*,y^*,t^*)$ of $P$, a pair of
optimal solutions to the LPs $(\ref{P})$ and $(\ref{D})$ is
$(x^*\alpha,y^*\alpha)$, or 
\rmitem{(b)}
the value of $P$ is positive, and then any max-min
strategy $(x,y,t)$ of $P$ fulfills 
$A x\le\0$,
$x\ge\0$,
$A\T y\ge\0$,
$y\ge\0$,
and
$c\T x>b\T y$, which shows that at least one LP is
infeasible.
\end{theorem}

The main effect of the definition of $P$ is that for any
min-max strategy $(x^*,y^*,t^*)$ with min-max value $v$,
we have
\begin{equation}
\label{mmBR}
\arraycolsep.2em
\begin{array}{rcrcrl}
&-&\alpha A\T y^*& \le&-c+\1v\\
\alpha Ax^*&&&\le&b+\1v\\
-\alpha c\T x^*&+&\alpha b\T y^*&\le&v\\
\end{array} 
\end{equation}
with \textit{constant} right-hand sides $-c$ and $b$
rather than these being scaled by $t^*$.
%
The number $\alpha$ is similar to the bound $M$ in
Theorem~\ref{t-DM} and (\ref{M}), because if the LPs
(\ref{P}) and (\ref{D}) have feasible solutions, then with
$x^*$ and $y^*$ as in Theorem~\ref{t-BR}(a), they have
feasible solutions $x^*\alpha,y^*\alpha$ with
$\1\T x^*\alpha+\1\T y^*\alpha \le \alpha$,
as noted by Brooks and Reny \cite[Remark~7]{BReny}.
If the value of $P$ in (\ref{PP}) is positive, then
any max-min strategy $(x,y,t)$ in Theorem~\ref{t-BR}(b)
proves the infeasibility of at least one of the LPs just as
in (\ref{witn}) in Theorem~\ref{t-BM}.

Given the constraints (\ref{mmBR}), the definition of $P$
can be seen as ``canonical'' as claimed by Brooks and
Reny, although one could also call it ``proof-induced''.
From the viewpoint of using this game, it has the 
disadvantage that all entries of $A$ are multiplied by the
large number $\alpha$, and $P$ is a full matrix and no
longer half-empty, with zero entries replaced by the rows of
$-c$ and $b$.
In contrast, in our matrix $B_M$ in Theorem~\ref{t-BM} the
large number $M$  appears in a single place, and the zero
entries remain.
The game $B_M$ also naturally extends Dantzig's original
game.

In summary, it seems that proving LP duality from the
minimax theorem requires quite a bit of linear algebra, most
concisely in our relatively short proof of
Theorem~\ref{t-gotu}.
We show an elegant use of linear algebra in the next,
final section.

\section{Minimally infeasible sets of inequalities}
\label{s-minfeas}

We conclude this article with a short elementary proof of
the Lemma of Farkas in its inequality-only version
(\ref{farkine}) due to
Conforti, Di Summa, and Zambelli \cite{conforti2007}.
The main trick is to state the \textit{minimal}
infeasibility of these inequalities in terms of
infeasibility of the corresponding equalities, which 
is canonically proved by induction.
The second step is to apply the linear algebra
Lemma~\ref{l-alt} to the infeasible equalities to obtain the
required vector~$y$ in (\ref{farkine}).

A set of linear equations and inequalities is called
infeasible if it has no solution, and \textit{minimally}
infeasible if omitting any one equation or inequality makes
it feasible.
The following proofs of theorem~2.1 and lemma~2.1 of
\cite{conforti2007}, in simplified notation,
show (\ref{farkine}) based on
minimally infeasible sets of inequalities.

\begin{theorem}[Conforti, Di Summa, and Zambelli \cite{conforti2007}]
\label{t-minfeas}
Let $A\in\reals^{m\times n}$ and $b\in\reals^n$
and let $a_1,\ldots,a_m$ be the rows of $A$.
Suppose the system $Ax\le b$ is minimally infeasible.
\rmitem{(i)}
Then the system $Ax=b$ is minimally infeasible.
\rmitem{(ii)}
Reversing any inequality $a_ix\le b_i$ 
in $Ax\le b$ creates a feasible system:
\begin{equation}
\label{xi}
\forall\,i\in[m]~~
\exists\,x^{(i)}\in\reals^n~:~
a_ix^{(i)}>b_i\,,~~~
\forall\,k\in[m]-\{i\}~:~
a_kx^{(i)}=b_k\,.
\end{equation}
\end{theorem}

\myproof
We prove that for any $R\subseteq[m]$
the constraints
\begin{equation}
\label{minfeas}
a_i x=b_i
\quad(i\in R),
\qquad
a_i x\le b_i
\quad(i\in [m]-R)
\end{equation}
are minimally infeasible.
The proof is by induction on $|R|$.
For $|R|=0$ condition (\ref{minfeas}) 
holds by assumption.
Suppose it holds for all $R$ up to a certain size $|R|$.
If $R=[m]$ then the proof of~(i) is complete, so let
$h\not\in R$, where we want to show that
\begin{equation}
\label{indfeas}
a_i x=b_i
\quad(i\in R),
\qquad
a_h x=b_h\,,
\qquad
a_i x\le b_i
\quad(i\in [m]-R-\{h\})
\end{equation}
is minimally infeasible.
The system (\ref{indfeas}) is infeasible (because $Ax\le b$
is infeasible), so we have to prove that omitting any
constraint $a_jx=b_j$ or $a_jx\le b_j$ for $j\in[m]$
produces a feasible system.
This is clearly the case if $j=h$, or if $j\in R$ by
applying the inductive hypothesis to $R\cup\{h\}-\{j\}$,
so let $j\not\in R$.
The constraints (\ref{minfeas}) for $i\ne h$ and $i\ne j$
have solutions $x^{(h)}$ and  $x^{(j)}$, respectively,
with 
\arraycolsep.2em
\begin{equation}
\label{jk}
\begin{array}{lcllclrcl}
a_i x^{(h)}&=&b_i
\quad(i\in R),&
\quad
a_i x^{(h)}&\le&b_i
\quad(i\in [m]-R-\{h\}),\quad
&
a_h x^{(h)}&>&b_h
\\ 
a_i x^{(j)}&=&b_i
\quad(i\in R),&
\quad
a_i x^{(j)}&\le&b_i
\quad(i\in [m]-R-\{j\}).
\\ 
\end{array}
\end{equation}
If $a_hx^{(j)}=b_h$ then $x^{(j)}$ is a feasible solution
to (\ref{indfeas}) with row $a_jx\le b_j$ omitted.
Otherwise $a_hx^{(j)}<b_h$\,, and a suitable convex
combination of $x^{(j)}$ and $x^{(h)}$ is such a solution
because $a_hx^{(h)}>b_h$.
This completes the induction.

Condition (ii) is an immediate consequence of (i):
Let $i\in[m]$.
Because $Ax=b$ is minimally infeasible, there is some
$x^{(i)}\in\reals^n$ such that
$a_kx^{(i)}=b_k$ for all $k\ne i$ and
$a_ix^{(i)}\ne b_i\,$,
where $a_ix^{(i)}<b_i$ would imply that $Ax\le b$ is
feasible, hence $a_ix^{(i)}> b_i\,$.
\endproof

\myproofof{$(\ref{farkine})$ using Theorem~\ref{t-minfeas}}
The direction ``$\Leftarrow$'' in (\ref{farkine}) is
immediate.
To prove ``$\Rightarrow$'', assume that $Ax\le b$ is
infeasible, and (by dropping sufficiently many rows from
these inequalities, whose components of~$y$ will be set to
zero) that $Ax\le b$ is minimally infeasible.
Denote the number of rows of this minimally infeasible
system again by~$m$.
By Theorem~\ref{t-minfeas}, $Ax=b$ is minimally infeasible.
By Lemma~\ref{l-alt}, there is some $y\in\reals^m$
such that 
$y\T[A~{-b}]=[\0\T~1]$.
We show that $y>\0$ (Adler, personal communication, 2023).
Let $i\in[m]$ and $x^{(i)}$ as in (\ref{xi}) in Theorem~\ref{t-minfeas}(ii).
Then
\[
1=[\0\T~1]\twovec{x^{(i)}}{1}=y\T[A~{-b}]\twovec{x^{(i)}}{1}=y_iz_i
\]
by (\ref{xi}) where $z_i=a_ix^{(i)}-b_i>0$, and thus $y_i>0$.
Hence, $y>\0$ as claimed.
\endproof

The proof of Theorem~\ref{t-minfeas} is canonical and 
easy to reconstruct.
As for proving the Lemma of Farkas,
in the same version (\ref{farkine}),
perhaps the most natural and elementary proof is
``projection'' or Fourier-Motzkin elimination (see
Schrijver \cite[p.~155f]{Schrijver} and references).
It expresses the constraints in $Ax\le b$ in terms of~$x_1$
by dividing each row by the coefficient of $x_1$ when it is
nonzero, which reverses the inequality when the coefficient
is negative.
This induces mutual bounds among the other linear
terms in $x_2,\ldots,x_n$ and eliminates~$x_1$.
This elimination is then iterated (and may lead to an
exponential increase in the number of constraints).
See Kuhn \cite{kuhn1956} and Tao \cite[p.~180]{tao2008} for
deriving (\ref{farkine}) in this way.

\section*{Acknowledgments}

I thank Ahmad Abdi, Ben Brooks, Phil Reny,
Giacomo Zambelli, and anonymous referees for helpful
comments and discussions,
and Sylvain Sorin for alerting me to the works of Loomis
\cite{loomis1946} and Ville \cite{ville1938}.
I thank Ilan Adler for great help in proving
Theorem~\ref{t-DM} without assuming strong LP duality.

\bibliographystyle{numeric}
\small 
\bibitemsep .4ex minus.1ex
\bibliography{bib-0sum}

\end{document}